\documentclass[pra,aps,twocolumn,superscriptaddress, longbibliography, notitlepage]{revtex4-1}
\usepackage[colorlinks=true, citecolor=blue, urlcolor=blue ]{hyperref}
\usepackage{graphicx}
\usepackage{dcolumn}
\usepackage{bm}
\usepackage{amsmath,amsfonts}
\usepackage[english]{babel}
\usepackage{amsthm}
\usepackage{hyperref}
\usepackage{xcolor}
\theoremstyle{plain}

\newtheorem{theorem}{Theorem}

\begin{document}
\title{Sharing tripartite nonlocality sequentially by arbitrarily many independent observers}

\author{Ya Xi}
\affiliation{Graduate School of China Academy of Engineering Physics, Beijing 100193, China}

\author{Mao-Sheng Li}
\affiliation{Department of Mathematics, South China University of Technology, GuangZhou 510640, China}

\author{Libin Fu}
\email{lbfu@gscaep.ac.cn}
\affiliation{Graduate School of China Academy of Engineering Physics, Beijing 100193, China}

\author{Zhu-Jun Zheng}
\affiliation{Department of Mathematics, South China University of Technology, GuangZhou 510640, China}

\begin{abstract}
	 There exist bipartite entangled states whose violations of Clauser-Horne-Shimony-Holt (CHSH) Bell inequality can be observed by  a single Alice and   arbitrarily many sequential Bobs [\href{https://link.aps.org/doi/10.1103/PhysRevLett.125.090401}{Phys. Rev. Lett. \textbf{125}, 090401 (2020)}].  Here we consider its analogues for tripartite systems: a tripartite entangled state is shared among Alice, Bob and multiple Charlies. The first Charlie measures his qubit and then passes his qubit to the next Charlie who measures again with other measurements and so on.   The goal is to maximize  the number of Charlies that can observe some kind of  nonlocality with the single Alice and Bob.  It has been shown that at most two Charlies could share genuine nonlocality of the Greenberger-Horne-Zeilinger (GHZ) state via the violation of Svetlichny inequality  with Alice and Bob [\href{https://doi.org/10.1007/s11128-018-2161-x}{Quantum Inf. Process. \textbf{18}, 42 (2019)} and \href{https://link.aps.org/doi/10.1103/PhysRevA.103.032216}{Phys. Rev. A \textbf{103}, 032216 (2021)}].   In this work, we show that arbitrarily many Charlies can have standard  nonlocality (via violations of  Mermin inequality) and some other kind of genuine nonlocality  (which is known as genuinely nonsignal nonlocality)  with the single Alice and single Bob.
	\end{abstract}
\maketitle

\section{Introduction}
Quantum nonlocality is one of the most striking features of quantum physics. Through the quantum violation of a suitable set of inequalities, Bell \cite{B64} demonstrated that the predictions of quantum mechanics are in contradiction with the classical causal relations. Beyond its importance in quantum foundations, quantum nonlocality is also the key resource for device-independent quantum information processing, such as building quantum protocols to decrease communication complexity \cite{dcc,dcc1} and providing secure quantum communication \cite{scc1,scc2}.

Recently, the question of whether a single entangled pair can generate a long sequence of nonlocal correlations has gained  extensive attentions. For the bipartite quantum systems, in \cite{Silva15,Mal16}, the authors show that at most two Bobs can achieve an expected CHSH violations with a single Alice when each Bob performs different measurements with equal probability, sharpness and the measurements of each Bob are independent of the choices of measurement settings and outcomes of the previous Bobs. In \cite{Brown20}, with unequal sharpness for each Bob's measurments, the authors prove that arbitrarily many independent Bobs can share the nonlocality of the Bell state with a single Alice. This result was soon extended to higher dimensional bipartite systems (see Ref. \cite{ZF21}).  There are also some progresses on the setting where multiple Alices and Bobs are considered \cite{Cheng21,Cheng22}.

It is natural to ask whether similar property holds in the multipartite setting.    For the tripartite quantum systems,   the authors  \cite{Saha19} showed  that via the violation of Mermin inequality, at most six Charlies can simultaneously demonstrate standard tripartite nonlocality with single Alice and single Bob. On the other hand, based on Svetlichny inequality \cite{S1987}, at most two Charlies can  simultaneously share genuine tripartite nonlocality with single Alice and single Bob \cite{Saha19,ZF21}.  As quantum nonlocality can be observed by violating different kinds of inequalities, it is interesting to ask whether there is some kind of nonlocality that can be detected sequentially by arbitrarily many Charlies with single Alice and single Bob via some related inequality. We will give  affirmed answers for the settings where  the nonlocality can be obtained via the violation of  Mermin inequality or the genuinely nonsignal nonlocality defined   in  Ref. \cite{Nonlocal2013}.  Contrary to the result that at most six Charlies can simultaneously observe the violations of Mermin inequality with single Alice and single Bob,  here we obtain that arbitrarily many independent Charlies can observe this violation with the single Alice and single Bob. This could be possible as we choose the different measurement strategy for Alice, Bob and multiple Charlies from the one in Ref. \cite{Saha19}.

The rest of this article is organized as follows. In Sec. \ref{sec:nonlocality}, we review  some definitions of the tripartite nonlocality.
  In Sec. \ref{sec:Scenario}, we will introduce the scenario of sharing tripartite nonlocality sequentially with multiple Charlies and single Alice and single Bob.
 In Sec. \ref{sec:Mermin}, we will give a constructive measurement strategy which enables that arbitrarily many independent Charlies can observe the violation of Mermin inquality with single Alice and Bob.    In Sec. \ref{sec:nonsignal}, we will provide a specific measurement strategy which enables  arbitrarily many independent Charlies to observe the genuinely nonsignal nonlocality  with  a single Alice and Bob.     Finally, we draw a conclusion  in section \ref{sec:conclusion}.

\section{Tripartite quantum nonlocality}\label{sec:nonlocality}
Different from the ones in bipartite systems, quantum states in tripartite systems can be not only entangled or nonlocally correlated, but also genuinely entangled or genuinely nonlocal correlated. Quantum nonlocality can be revealed via violations of various Bell inequalities.
For the tripartite quantum systems, except for the well-known Svetlichny inequalities \cite{S1987},
in \cite{Nonlocal2013} other three-qubit genuine nonlocality and 3-way nonlocal correlations, have been studied.

Now we consider a tripartite scenario where each of three spatially separated parties, Alice, Bob and Charlie performs the measurements $X_{i}, Y_{j}, Z_{k}$ on their subsystems, respectively with
outcomes $A$, $B$ and $C$, $i,j,k\in\{0, 1\}, A,B,C\in\{0, 1\}$. Let $P(ABC|X_{i}Y_{j}Z_{k})$ denote the joint outcome probabilities where Alice measures her system by $X_i$ with outcome $A$ (similar for Bob and Charlie). First, if the probability correlations $P(ABC|X_{i}Y_{j}Z_{k})$ among the measurement outcomes can be written as
\begin{equation*}
  P(ABC|X_{i}Y_{j}Z_{k})=\sum\limits_{\lambda}q_{\lambda}P_{\lambda}(A|X_{i})P_{\lambda}(B|Y_{j})P_{\lambda}(C|Z_{k}),
\end{equation*}
with $0\leq q_{\lambda}\leq1$ and $\sum_{\lambda}q_{\lambda}=1$, then it is called fully local. If $P(ABC|X_{i}Y_{j}Z_{k})$ is not fully local, we call $P(ABC|X_{i}Y_{j}Z_{k})$ exhibits \emph{standard tripartite nonlocality}. In particular, it can be detected by the violations of the Mermin inequalities \cite{Mermin90}, which have the following form:
\begin{equation}\label{Mermin equality}
 \langle X_{1}Y_{0}Z_{0}\rangle+\langle X_{0}Y_{1}Z_{0}\rangle+\langle X_{0}Y_{0}Z_{1}\rangle-\langle X_{1}Y_{1}Z_{1}\rangle\leq2,
\end{equation}
where $\langle X_{i}Y_{j}Z_{k}\rangle=\sum_{ABC}(-1)^{A+B+C}P(ABC|X_{i}Y_{j}Z_{k})$.
As  it is pointed out by Svetlichny \cite{S1987}, if the correlation can be written as the form
\begin{equation}\label{eq:S_2}
	\begin{array}{ccl}
	P(ABC|X_{i}Y_{j}Z_{k})   & = & \displaystyle\sum_{\lambda}q_{\lambda} P_{\lambda}(AB|X_{i}Y_{j}) P_{\lambda}(C|Z_{k})  \\[2mm]
	& + &  \displaystyle\sum_{\mu}q_{\mu} P_{\mu}(AC|X_{i}Z_{k}) P_{\mu}(B|Y_{j})  \\[2mm]
	& + &  \displaystyle\sum_{\nu}q_{\nu} P_{\nu} (BC|Y_{j}Z_{k})P_{\nu} (A|X_{i}),
\end{array}
\end{equation}
where $0\leq q_{\lambda}$, $q_{\mu}, q_{\nu} \leq 1,$  $\sum_{\lambda}q_{\lambda}+\sum_{\mu}q_{\mu}+\sum_{\nu}q_{\nu}=1$, then $P(ABC|X_{i}Y_{j}Z_{k})$ is called $S_2$ local.  Otherwise, it is called \emph{$3$-way genuine nonlocality} which is also known as Svetlichny nonlocality.   Svetlichny  found that  three-way genuine nonlocality can be  observed by violations of  Svetlichny  inequality which is defined as
\begin{equation*}\label{eq:Svetlichy_equality}
\begin{array}{rl}
		&\langle X_{0}Y_{0}Z_{0}\rangle+ \langle X_{0}Y_{1}Z_{0}\rangle+\langle X_{0}Y_{0}Z_{1}\rangle-\langle X_{0}Y_{1}Z_{1}\rangle\\[2mm]
		+& \langle X_{1}Y_{0}Z_{0}\rangle-\langle X_{1}Y_{1}Z_{0}\rangle-\langle X_{1}Y_{0}Z_{1}\rangle+\langle X_{1}Y_{1}Z_{1}\rangle\leq 4.
		\end{array}
\end{equation*}

In particular, even if $P(ABC|X_{i}Y_{j}Z_{k})$ violates the Mermin equality, it does not necessarily demonstrate that the correlation exhibits  Svetlichny nonlocality. In \cite{Nonlocal2013}, the authors introduced two alternative definitions of three-way nonlocality, strictly weaker than Svetlichny nonlocality. Here apart from Svetlichny nonlocality, we mainly study the definition $1$ in \cite{Nonlocal2013} about the genuine nonlocality.

We assume that the probabilities $P(ABC|X_{i}Y_{j}Z_{k})$  satisfy Eq. \eqref{eq:S_2}. Moreover,   for any possible
$ A, B, C, C^{\prime}$,  and  $X_{i}, Y_{j}, Z_{k}, Z^{\prime}_{k}$,  the equalities $$\sum\limits_{B} P_\lambda(AB|X_{i}Y_{j})=\sum\limits_{B^{\prime}}P_\lambda(AB^{\prime}|X_{i}Y^{\prime}_{j})$$ are satisfied and other equalities can be also obtained from permutations of the parties, then the correlations are called nonsignal-local (denoted $NS$ local). Otherwise we call them \emph{genuinely nonsignal nonlocality} (i.e., genuinely $NS$ nonlocality).

To detect the $NS$ genuine nonlocality, we consider the following inequality (denoted $NS$ inequality):
\begin{equation}\label{NS_inequality}
	\langle Y_{0}Z_{0}\rangle+\langle X_{0}Z_{0}\rangle+\langle X_{1} Y_{0}\rangle-\langle X_{0}Y_{1}Z_{1}\rangle+\langle X_{1}Y_{1}Z_{1}\rangle\leq3,
\end{equation}
where $\langle X_{i}Y_{j}\rangle=\sum_{AB}(-1)^{A+B}P(AB|X_{i}Y_{j})$.

\begin{figure}[ptb]
	\includegraphics[width=0.5\textwidth]{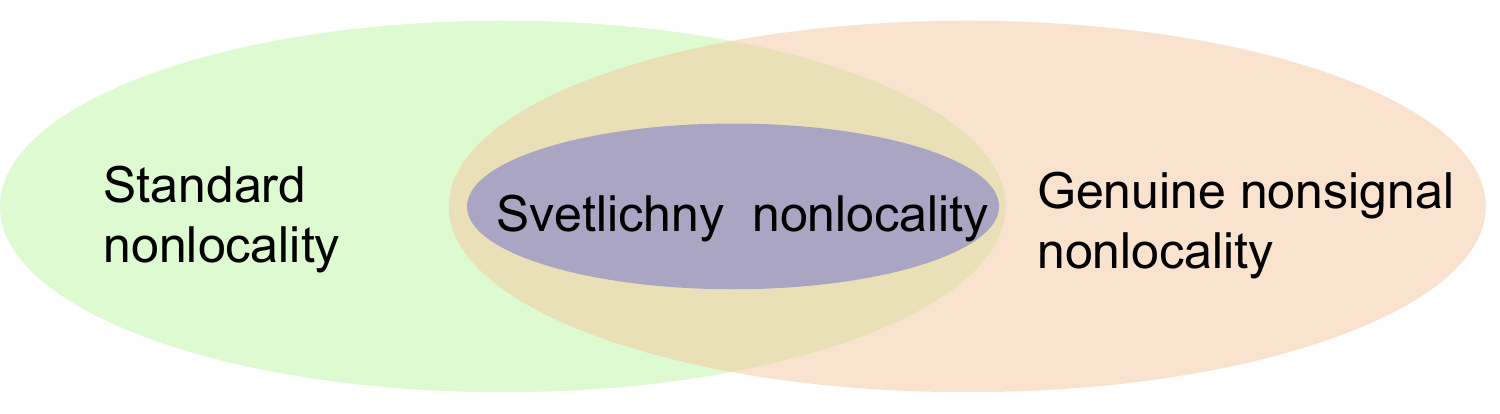}
	\caption{This figure shows the relations among standard nonlocality, Svetlichny nonlocality  and genuinely nonsignal nonlocality.}
	\label{fig:relation}%
\end{figure}

By definitions, the relations among standard nonlocality, Svetlichny nonlocality  and genuinely nonsignal nonlocality can be seen in Fig. \ref{fig:relation}.

\section{Scenario of sharing of tripartite nonlocality by multiple Charlies }\label{sec:Scenario}
Denote  $\sigma_{i},$ for $ i\in \{1,2,3\}$  to be  the Pauli operators.
Throughout this paper, we use two-outcome positive operator-valued measurements (POVMs), $\{E, \mathbb{I}-E\}$, where $E$ has the form $E=\frac{\mathbb{I}+\gamma\sigma_{\vec{r}}}{2}$, $\gamma\in[0,1]$ is the sharpness of the measurement, $\vec{r}=(r_1,r_2,r_3)\in\mathbb{R}^{3}$, $\parallel\vec{r} \parallel=1$, $\sigma_{\vec{r}}=r_{1}\sigma_{1}+r_{2}\sigma_{2}+r_{3}\sigma_{3}.$  For example, $\{A_{0|0}, A_{1|0} \}$ where  $A_{0|0}=\frac{1}{2}(\mathbb{I} +\sigma_1)$ and $A_{1|0}=\mathbb{I}-A_{0|0}=\frac{1}{2}(\mathbb{I} -\sigma_1).$ Therefore, to define a two-outcome measurement, it is enough to define one measurement element.

\begin{figure*}[ptb]
	\includegraphics[width=0.7\textwidth]{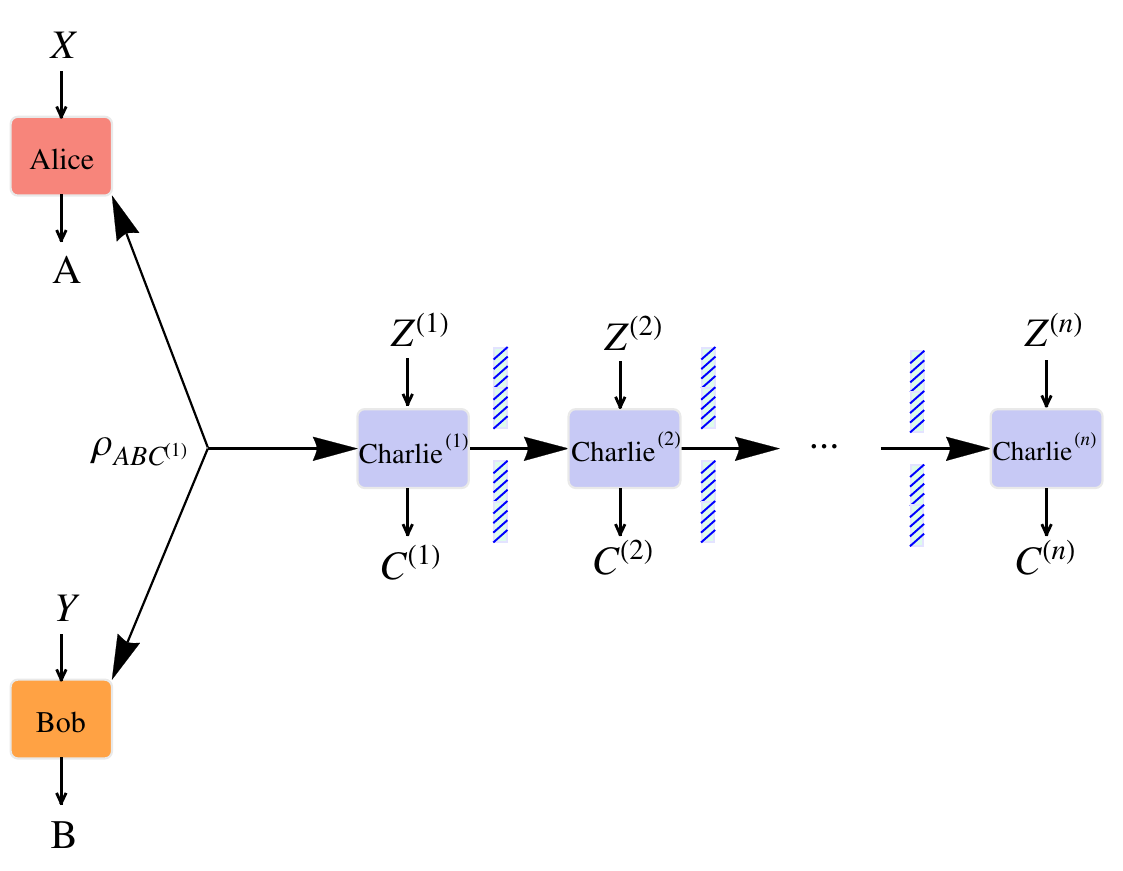}
	\caption{Sharing the genuine tripartite nonlocality with multiple Charlies: A quantum state $\rho_{ABC^{(1)}}$ is initially distributed between Alice, Bob and Charlie$^{(1)}$. After Charlie$^{(1)}$ performs his randomly selected measurement and records the outcomes, he passes the post-measurement quantum state to Charlie$^{(2)}$ who then repeats the process. In particular, the measurement choices and outcomes of each Charlie are not conveyed.}
	\label{Fig1}%
\end{figure*}

Now we introduce the scenario of sharing tripartite nonlocality sequentially with multiple Charlies and single Alice and single Bob.  The corresponding measurement scenario illustrated in Fig. \ref{Fig1} is considered.

Three particles are prepared in the state $\rho_{ABC^{(1)}}=|\mathrm{GHZ}\rangle\langle \mathrm{GHZ}|$ where $|\mathrm{GHZ}\rangle=\frac{1}{\sqrt{2}}(|000\rangle+|111\rangle).$   These three particles are spatially separated and shared between Alice, Bob and multiple Charlies (i.e., Charlie$^{(1)}$, Charlie$^{(2)}$, Charlie$^{(3)}$,..., Charlie$^{(n)}$). Alice performs the measurement $X$ on the first particle and gets the outcome $A$. Bob performs the measurement $Y$ on the second particle and gets the outcome $B$. And multiple Charlies perform the measurements $Z^{(k)}$on the third particle and get the outcomes $C^{(k)}$ sequentially. In particular, Charlie$^{(1)}$ performs measurements on the third particle and after doing measurements, he passes the particle to Charlie$^{(2)}$. Then Charlie$^{(2)}$ also delivers the particle to Charlie$^{(3)}$ after doing measurements and so on.
Moreover, each Charlie performs measurements independent of the measurement choices and outcomes of the previous Charlies on this sequence. And here we consider the unbiased input scenario, i.e., all possible measurement settings of each Charlie are uniformly distributed.

 The goal is to maximize  the number of Charlies that can observe some kind of  nonlocality with single Alice and Bob.  Therefore, it is crucial to find out what the state $\rho_{ABC^{(k)}}$  shared by Alice, Bob, and $\mathrm{Charlie}^{(k)}$ is after the $\mathrm{Charlie}^{(k-1)}$ performs his measurements.  In fact, suppose Charlie$^{(k-1)}$ performed the measurement according to $Z^{(k-1)}=z$ and received the outcome $C^{(k-1)}=c$, the post-measurement state can be described by the L\"{u}ders rule
 \begin{equation}\label{eq:state_recursive}
 	\rho_{ABC^{(k)}}=\frac{1}{2}\sum\limits_{c,z}(\mathbb{I}\otimes\mathbb{I}\otimes\sqrt{C^{(k-1)}_{c|z}}\rho_{ABC^{(k-1)}}\mathbb{I}\otimes\mathbb{I}\otimes\sqrt{C^{(k-1)}_{c|z}}).
 \end{equation}

\section{Sharing of tripartite nonlocality by multiple Charlies via Mermin Inequality}\label{sec:Mermin}
First, we consider how many Charlies can simultaneously demonstrate tripartite nonlocality via Mermin inequality \eqref{Mermin equality} with single Alice and Bob.  Therefore, it is important to find out the Mermin value  $\mathbf{I}^{(k)}_{M}$ among Alice, Bob and Charlie$^{(k)}$ which is defined by
\begin{equation}\label{Mermin Value}
  \mathrm{Tr}[\rho_{ABC^{(k)}}(X_{1}Y_{0}Z^{(k)}_{0}+X_{0}Y_{1}Z^{(k)}_{0}+X_{0}Y_{0}Z^{(k)}_{1}-X_{1}Y_{1}Z^{(k)}_{1})].
\end{equation}
Here $X_0,X_1,Y_0,Y_1,Z^{(k)}_0,Z^{(k)}_1$ are the  observables  corresponding to their measurements which will be defined as  follows.

To explain how we can define a sequence of pairs of POVMs for Alice, Bob, Charlie$^{(k)}$, such that $\mathbf{I}^{(k)}_{M}>2$, $k\in\{1,2,\cdots,n\}$, we give the following measurement strategy about Alice, Bob and Charlie$^{(k)}$.
In this measurement strategy, Alice's POVMs are defined by
\begin{eqnarray}\label{eq:A1}
  A_{0|0} =\frac{\mathbb{I}+\sigma_{1}}{2}, \ \
  A_{0|1} = \frac{\mathbb{I}+\sigma_{2}}{2},
\end{eqnarray}
Bob's POVMs are defined by
\begin{eqnarray}\label{eq:B1}
  B_{0|0} = \frac{\mathbb{I}-\theta\sigma_{2}}{2}, \ \
  B_{0|1} = \frac{\mathbb{I}+\theta\sigma_{1}}{2},
\end{eqnarray}
for $\theta\in(0,1)$. For each $k=1,2,\cdots,n$,  Charlie$^{(k)}$'s POVMs are defined by
\begin{eqnarray}\label{eq:C1}
  C_{0|0}^{(k)} = \frac{\mathbb{I}+\sigma_{1}}{2}, \ \
  C_{0|1}^{(k)} = \frac{\mathbb{I}+\gamma_{k}\sigma_{2}}{2},
\end{eqnarray}

So the observables are given by $X_{i}= A_{0|i}-A_{1|i}$, $Y_{i}= B_{0|i}-B_{1|i}$, $Z^{(k)}_{i}= C^{(k)}_{0|i}-C^{(k)}_{1|i}$, $i=0,1.$ Under these measurements  and   the initial state $|\mathrm{GHZ}\rangle=\frac{1}{\sqrt{2}}(|000\rangle+|111\rangle),$    we can calculate out the expected Mermin value of Alice, Bob and Charlie$^{(k)}$ as follows   (see Appendix \ref{appen:I_m} for the detailed calculation)
\begin{equation}\label{eq:I_m}
\mathbf{I}^{(k)}_{M}=2^{2-k}\theta[\gamma_{k}+\prod\limits_{j=1}^{k-1}(1+\sqrt{1-\gamma_{j}^{2}})].
\end{equation}
The inequality $\mathbf{I}^{(k)}_{M}>2$ implies Alice, Bob and Charlie$^{(k)}$ can observe the standard nonlocality. To ensure arbitrarily  many Charlies can share the standard nonlocality with single Alice and Bob, it is suffcient to prove that for any $n\in  \mathbb{N}$,  there exists some $(\theta, \gamma_{k})$ such that  $\mathbf{I}^{(k)}_{M}>2$ holds for all $k=1,2,\cdots,n$. By Eq. \eqref{eq:I_m}, we have
\begin{equation}\label{eq:Mermineq}
 \mathbf{I}^{(k)}_{M}>2\Leftrightarrow \gamma_{k}>\frac{2^{k-1}}{\theta}-\prod\limits_{j=1}^{k-1}(1+\sqrt{1-\gamma_{j}^{2}})
\end{equation}
which motivates us  to find a sequence $\{\gamma_{k}(\theta)\}$ such that for $\forall k \in \{1,2,\cdots,n\},$ $\gamma_{k}(\theta)\in[0,1]$ and $\gamma_{k}>\frac{2^{k-1}}{\theta}-\prod\limits_{j=1}^{k-1}(1+\sqrt{1-\gamma_{j}^{2}}).$

To achieve this,
 we will give a specific sequence and prove that this sequence will satisfy the above conditions.  In fact,   set $\epsilon>0$,  and $\gamma_{1}(\theta):=
(1+\epsilon)(\frac{1}{\theta}-1)$ and for $k\geq 2$
\begin{equation}\label{}
  \gamma_{k}(\theta)=\left\{
                       \begin{array}{ll}
                       (1+\epsilon)(\frac{2^{k-1}}{\theta}-P_{k}) , & \hbox{$0\leq\gamma_{k-1}(\theta)\leq1$ ;} \\
                         \infty, & \hbox{others}
                       \end{array}
                     \right.
\end{equation}
where $P_{k}=\prod\limits_{j=1}^{k-1}(1+\sqrt{1-\gamma_{j}^{2}}).$ Then we have the following statement which is sufficient to deduce that  arbitrarily  many Charlies can share the standard nonlocality of $\rho_{ABC^{(1)}}$ with single Alice and Bob.

\begin{theorem}\label{theorem2}
For each $n\in\mathbb{N},$ there exists a sequence $\{\gamma_{k}(\theta)\}_{k=1}^{n}$ and $\theta_{n}\in(0,1)$ such that $\mathbf{I}^{(k)}_{M}>2$ and $0<\gamma_{k}(\theta)<1, \theta\in(\theta_{n}, 1),$ for $k=1,2,\cdots, n$.
\end{theorem}
The proofs of   Theorem \ref{theorem2} are given in Appendix \ref{appen:TheoremI_m}. Theorem \ref{theorem2} shows that by initially sharing the $\mathrm{GHZ}$ state $|\mathrm{GHZ}\rangle=\frac{1}{\sqrt{2}}(|000\rangle+|111\rangle)$, the numbers of Charlies that violate the Mermin inequality with single Alice and Bob are unbounded.

One notes that the authors \cite{Saha19} showed that at most six Charlies can simultaneously demonstrate standard tripartite nonlocality with single Alice and single Bob where they assumed that the measurements of both Alice and Bob are projective measurements. In the contradiction with the results from \cite{Saha19}, by  Theorem \ref{theorem2} and the following {\it Remark 1}, we show that using the different measurement strategies for Alice, Bob and Charlies enables more Charlies to share the standard tripartite nonlocality.

{\it Remark 1.} 1) We state that  when
the initial state is $|\mathrm{GHZ}\rangle=\frac{1}{\sqrt{2}}(|000\rangle+|111\rangle),$ even if both Alice and Bob perform the projective measurements, i.e. $\theta=1,$ there also exist unbounded numbers of Charlies such that $\rho_{ABC^{(k)}}$ can violate the Mermin inequality with single Alice and Bob. However, the  measurement stategies defined by Eqs. \eqref{eq:A1}, \eqref{eq:B1}, and \eqref{eq:C1} should be slightly changed as follows,  $ \gamma_1(\varphi)=(1+\epsilon)\varphi$ and  for $k\geq 2$
 \begin{equation}\label{}
 	\gamma_{k}(\varphi)=\left\{
 	\begin{array}{ll}

 		(1+\epsilon)( 2^{k-1} -P_{k}(\varphi)) , &  \hbox{$0\leq\gamma_{k-1}(\varphi)\leq1$ ;} \\
 		\infty, & \hbox{others}
 	\end{array}
 	\right.
 \end{equation}
 where $P_{k}(\varphi)=\prod\limits_{j=1}^{k-1}(1+\sqrt{1-\gamma_{j}^{2}(\varphi)}).$ In fact, in the case of $\theta=1$,  Eq. \eqref{eq:Mermineq} could be changed into
 \begin{equation}\label{eq:Mermineqd}
 	\mathbf{I}^{(k)}_{M}>2\Leftrightarrow \gamma_{k}> 2^{k-1} -\prod\limits_{j=1}^{k-1}(1+\sqrt{1-\gamma_{j}^{2}}).
 \end{equation}
Therefore, to ensure the first Charlie obtain a violation of Mermin inequality it is sufficient to make sure $0<\gamma_1<1$ which could be satisfied by choosing a small $\varphi>0$ in our setting. With similar arguments as  Appendix \ref{appen:TheoremI_m},   one could   deduce that $\lim\limits_{\varphi\rightarrow 0^{+}}\gamma_{k}(\varphi)=0$ for all $k$ which is sufficient to yield our statement. 

2) We demonstrate that if Alice performs the sharp measurements and one of Bob's measurements is sharp, then there also exists unbounded numbers of Charlies such that $\rho_{ABC^{(k)}}$ can violate the Mermin inequality for any times with single Alice and Bob when the initial state is $|\mathrm{GHZ}\rangle=\frac{1}{\sqrt{2}}(|000\rangle+|111\rangle)$ based on the following measurement strategy:

Alice's POVMs are defined by
\begin{eqnarray}
	A_{0|0}  =  \frac{\mathbb{I}+\sigma_{1}}{2}, \   \
	A_{0|1}  = \frac{\mathbb{I}+\sigma_{2}}{2}.
\end{eqnarray}
Bob's POVMs are defined by
\begin{eqnarray}
  B_{0|0} = \frac{\mathbb{I}-\theta_{1}\sigma_{2}}{2}, \ \
  B_{0|1} = \frac{\mathbb{I}+\theta_{2}\sigma_{1}}{2},
\end{eqnarray}
for $\theta_{1}, \theta_{2}\in[0,1].$ For each $k=1,2,\cdots,n$,  Charlie$^{(k)}$'s POVMs are defined by
\begin{eqnarray}
  C_{0|0}^{(k)} = \frac{\mathbb{I}+\sigma_{1}}{2}, \ \
  C_{0|1}^{(k)} = \frac{\mathbb{I}+\gamma_{k}\sigma_{2}}{2}.
\end{eqnarray}
Here we assume $\theta_{1}=1, \theta_{2}\neq1,$ under these measurements  and   the initial state $|\mathrm{GHZ}\rangle=\frac{1}{\sqrt{2}}(|000\rangle+|111\rangle),$ we can get
\begin{equation}
\mathbf{I}^{(k)}_{M}=2^{1-k}(1+\theta_{2})[\gamma_{k}+\prod\limits_{j=1}^{k-1}(1+\sqrt{1-\gamma_{j}^{2}})].
\end{equation}

In order to observe $\mathbf{I}^{(k)}_{M}>2$, we will need
\begin{equation}
 \mathbf{I}^{(k)}_{M}>2\Leftrightarrow \gamma_{k}>\frac{2^{k}}{1+\theta_{2}}-\prod\limits_{j=1}^{k-1}(1+\sqrt{1-\gamma_{j}^{2}}).
\end{equation}
Next we can define $\{\gamma_{k}(1,\theta_{2})\}$, for some fixed $\epsilon>0$,
\begin{equation}\label{}
  \gamma_{k}(1,\theta_{2})=\left\{
                       \begin{array}{ll}
                         (1+\epsilon)(\frac{2}{1+\theta_{2}}-1), & \hbox{$k=1$;} \\
                     (1+\epsilon)(\frac{2^{k}}{1+\theta_{2}}-P_{k}) , & \hbox{$0\leq\gamma_{k-1}(1,\theta_{2})\leq1$ ;} \\
                         \infty, & \hbox{others.}
                       \end{array}
                     \right.
\end{equation}
where $P_{k}=\prod\limits_{j=1}^{k-1}(1+\sqrt{1-\gamma_{j}^{2}}).$

With similar methods as  Appendix \ref{appen:TheoremI_m} in our paper,   one could   deduce that $\lim\limits_{\theta_{2}\rightarrow 1^{-}}\gamma_{k}(1, \theta_{2})=0$ for all $k$ and further demonstrate our statement.

{\it Remark 2.} When  starting with the $\mathrm{W}$ state $|\mathrm{W}\rangle=\frac{1}{\sqrt{3}}(|100\rangle+|010\rangle+|001\rangle)$, we prove that at most two Charlies can demonstrate standard nonlocality through the violation of Mermin inequality with the single Alice and Bob by the following measurement strategy:

Alice's POVMs are defined by
\begin{eqnarray}
	A_{0|0} &=& \frac{\mathbb{I}+\cos(\theta_{1})\sigma_{3}-\sin(\theta_{1})\sigma_{1}}{2}, \\
	A_{0|1} &=& \frac{\mathbb{I}+\sin(\theta_{1})\sigma_{3}+\cos(\theta_{1})\sigma_{1}}{2}.
\end{eqnarray}
Bob's POVMs are defined by
\begin{eqnarray}
	B_{0|0} &=& \frac{\mathbb{I}+\cos(\theta_{2})\sigma_{3}-\sin(\theta_{2})\sigma_{1}}{2}, \\
	B_{0|1} &=& \frac{\mathbb{I}+\sin(\theta_{2})\sigma_{3}+\cos(\theta_{2})\sigma_{1}}{2},
\end{eqnarray}
for $\theta_{i}\in[0,\frac{\pi}{2}], i\in\{1,2\}$. For each $k=1,2,\cdots,n$,  Charlie$^{(k)}$'s POVMs are defined by
\begin{eqnarray}
  C_{0|0}^{(k)} = \frac{\mathbb{I}-\sigma_{3}}{2}, \ \
  C_{0|1}^{(k)} = \frac{\mathbb{I}-\gamma_{k}\sigma_{1}}{2}.
\end{eqnarray}

The proof of {\it Remark 2} is given in Appendix \ref{appen:Remark 2}. Note that when $\theta_1+\theta_2=\frac{\pi}{2}$ and $\gamma_1=1$, Alice , Bob, and  the first Charlie could yield a maximal violation (with value $3$) of Mermin inequality under the assumption that the initial  state is  W state.  However, there may exist some other measurement strategy with a larger number of Charlies that could yield a  violation of Mermin inequality with single Alice and Bob.

\section{Sharing of genuine tripartite nonlocality by multiple Alices via NS inequality}\label{sec:nonsignal}
Second, we consider how many Charlies can simultaneously demonstrate tripartite genuinely nonsignal nonlocality through $NS$ inequality \eqref{NS_inequality} with a single Alice  and Bob. Therefore, we need to  calculate out the  $NS$ value between Alice, Bob and Charlie$^{(k)}$. The $NS$ value is defined as

\begin{multline}
 \mathbf{I}^{(k)}_{NS}\equiv \mathrm{Tr}[\rho_{ABC^{(k)}}( Y_{0}Z_{0}^{(k)}+X_{0}Z_{0}^{(k)}\\
 + X_{1}Y_{0}- X_{0}Y_{1}Z_{1}^{(k)}+X_{1}Y_{1}Z_{1}^{(k)})].
\end{multline}

To explain how we can define a sequence of pairs of POVMs for Alice, Bob, and Charlie$^{(k)}$ such that $\mathbf{I}^{(k)}_{NS}>3$, $k\in\{1,2,\cdots,n\}$, we give the following measurement strategy about Alice, Bob and Charlie$^{(k)}$.
In this measurement strategy, Alice's POVMs are defined by
\begin{eqnarray}
	A_{0|0} &=& \frac{\mathbb{I}+\cos(\theta)\sigma_{3}-\sin(\theta)\sigma_{1}}{2}, \\
	A_{0|1} &=& \frac{\mathbb{I}+\cos(\theta)\sigma_{3}+\sin(\theta)\sigma_{1}}{2},
\end{eqnarray}
for $\theta\in[0,\frac{\pi}{2}].$
Bob's POVMs are defined by
\begin{eqnarray}
  B_{0|0} =  \frac{\mathbb{I}+\sigma_{3}}{2}, \ \
  B_{0|1}  =  \frac{\mathbb{I}+\sigma_{1}}{2},
\end{eqnarray}
for each $k=1,2,\cdots,n$, Charlie$^{(k)}$'s POVMs are defined by
\begin{eqnarray}
	C_{0|0}^{(k)}= \frac{\mathbb{I}+\sigma_{3}}{2}, \  \
	C_{0|1}^{(k)} = \frac{\mathbb{I}+\gamma_{k}\sigma_{1}}{2},
\end{eqnarray}

The observables are given by $X_{i}= A_{0|i} -A_{1|i} $, $Y_{i}= B_{0|i}-B_{1|i}$, $Z_{i}^{(k)}= C_{0|i}^{(k)}-C_{1|i}^{(k)}$, $i=0,1.$ For these measurements and the initial state $|\mathrm{GHZ}\rangle=\frac{1}{\sqrt{2}}(|000\rangle+|111\rangle)$,   we can calculate out the expected  $NS$ value  of Alice, Bob and Charlie$^{(k)}$ as follows   (see Appendix \ref{appen:I_NS} for the detailed calculation)

\begin{equation}\label{eq:I_NS}
\mathbf{I}^{(k)}_{NS}=(\cos \theta +1)\frac{\prod\limits_{j=1}^{k-1}(1+\sqrt{1-\gamma_{j}^{2}})}{2^{k-1}}+\cos\theta+2^{2-k}\gamma_{k}\sin\theta.
\end{equation}
 The inequality $\mathbf{I}^{(k)}_{NS}>3$ implies Alice, Bob and Charlie$^{(k)}$ can observe the genuinely nonsignal  nonlocality. To ensure arbitrarily  many Charlies can share the genuinely nonsignal  nonlocality with single Alice and Bob, it is suffcient to prove that for any $n\in  \mathbb{N}$,  there exists some $\theta$ such that  $\mathbf{I}^{(k)}_{NS}>3$ holds for all $k=1,2,\cdots,n$. By Eq. \eqref{eq:I_NS}, we have
\begin{equation}\label{}
	\mathbf{I}^{(k)}_{NS}>3\Leftrightarrow \gamma_{k}>\frac{3-\cos\theta-(1+\cos\theta)\frac{\prod\limits_{j=1}^{k-1}(1+\sqrt{1-\gamma_{j}^{2}})}{2^{k-1}}}{2^{2-k}\sin\theta}
\end{equation}
 which motivates us  to find a sequence $\{\gamma_{k}(\theta)\}$ such that for $\forall k \in \{1,2,\cdots,n\},$ $\gamma_{k}(\theta)\in[0,1]$ and
  $$ \gamma_{k}>\frac{3-\cos(\theta)-(1+\cos(\theta))\frac{\prod\limits_{j=1}^{k-1}(1+\sqrt{1-\gamma_{j}^{2}})}{2^{k-1}}}{2^{2-k}\sin(\theta)}.$$

 To acheive this,
 we will give a specific sequence and prove that this sequence will satisfy the above conditions.  In fact,   set $\epsilon>0$,  and $\gamma_{1}(\theta):=
 (1+\epsilon)\frac{1-\cos(\theta)}{\sin(\theta)}$ and for $k\geq 2$
 \begin{equation}\label{}
 	\gamma_{k}(\theta)=\left\{
 	\begin{array}{ll}
 		(1+\epsilon)\frac{3-\cos(\theta)-(1+\cos(\theta))\frac{P_{k}}{2^{k-1}}}{2^{2-k}\sin(\theta)} , & \hbox{$0\leq\gamma_{k-1}(\theta)\leq1$ ;} \\
 		\infty, & \hbox{others,}
 	\end{array}
 	\right.
 \end{equation}
 where $P_{k}=\prod\limits_{j=1}^{k-1}(1+\sqrt{1-\gamma_{j}^{2}}).$

  Then we have the following statement which is sufficient to deduce that  arbitrarily  many Charlies can share the genuinely nonsignal nonlocality of $\rho_{ABC^{(1)}}$ with single Alice and Bob.

\begin{theorem}\label{theorem4}
For each $n\in\mathbb{N},$ there exists a sequence $\{\gamma_{k}(\theta)\}_{k=1}^{n}$ and a $\theta_{n}\in(0,1)$ such that $\mathbf{I}^{(k)}_{NS}>3$ and $0<\gamma_{k}(\theta)<1, \theta\in(0,\theta_{n}),$ for $k=1,2,\cdots, n$.
\end{theorem}
The proofs of Theorem \ref{theorem4} are given in Appendix \ref{appen:TheoremI_NS}. Theorem \ref{theorem4} shows that by initially sharing the maximally entangled state $|\mathrm{GHZ}\rangle=\frac{1}{\sqrt{2}}(|000\rangle+|111\rangle)$, the numbers of Charlies that violate the $NS$ inequality with a single Alice and Bob are unbounded.

{\it Remark 3.} When   starting with the $\mathrm{W}$ state $|\mathrm{W}\rangle=\frac{1}{\sqrt{3}}(|100\rangle+|010\rangle+|001\rangle)$, we prove that at most one Charlie can demonstrate the tripartite genuinely nonsignal nonlocality through $NS$ inequality with the single Alice and Bob by the following measurement strategy:

Alice's POVMs are defined by
\begin{eqnarray}
  A_{0|0} =\frac{\mathbb{I}+\sigma_{3}}{2}, \ \
  A_{0|1} = \frac{\mathbb{I}+\sigma_{1}}{2}.
\end{eqnarray}
Bob's POVMs are defined by
\begin{eqnarray}
  B_{0|0} = \frac{\mathbb{I}-\sigma_{3}}{2}, \ \
  B_{0|1} = \frac{\mathbb{I}+\sigma_{1}}{2}.
\end{eqnarray}
For each $k=1,2,\cdots,n$,  Charlie$^{(k)}$'s POVMs are defined by
\begin{eqnarray}
  C_{0|0}^{(k)} = \frac{\mathbb{I}+\sigma_{3}}{2}, \ \
  C_{0|1}^{(k)} = \frac{\mathbb{I}+\gamma_{k}\sigma_{1}}{2}.
\end{eqnarray}
Moreover, the corresponding $NS$ inequality is choosing to be:
\begin{equation}
	\langle X_{1}Y_{1}\rangle+\langle Y_{0}Z_{0}\rangle+\langle X_{1} Z_{1}\rangle+\langle X_{0}Y_{0}Z_{0}\rangle-\langle X_{1}Y_{0}Z_{1}\rangle\leq3.
\end{equation}
The proof of {\it Remark 3} is given in Appendix \ref{appen:Remark 3}.  Note that when $\gamma_1=1$, Alice , Bob, and  the first Charlie could yield a maximal violation (with value $\frac{10}{3}$) of NS inequality under the assumption that the initial  state is  W state.  However, there may exist some other measurement strategy with a larger number of Charlies that could yield a  violation of NS inequality with single Alice and Bob.

\section{ Conclusions and discussions}\label{sec:conclusion}
In this work, we consider the sequential detection of quantum standard and genuinely nonsignal  nonlocality in the tripartite quantum systems.  Just like the bipartite settings, using the different measurement strategies will enable more observers to share the tripartite nonlocality.  By this way, we deduce that arbitrarily many independent Charlies could observe the standard tripartite nonlocality  or genuinely nonsignal one of $|\mathrm{GHZ}\rangle$ with single Alice and Bob. Moreover, we also prove that when the initial state is $|\mathrm{W}\rangle$, for the standard nonlocality, at most two Charlies can share the nonlocality with single Alice and Bob through the Mermin inequality; for the genuinely nonsignal nonlocality, at most one Charlie can demonstrate this genuine nonlocality with single Alice and Bob by the $NS$ inequality.

One of the most important applications based on these nonlocal correlations in quantum protocols is the device-independent random number generation. The amount of randomness from the measurement outcomes on quantum systems is quantified by the guessing probability and can generally be bounded numerically or analytically. The quantitative relationship between nonlocality and maximum certifiable randomness is difficult to be exploited. In the bipartite systems \cite{A12}, for the standard Bell scenario where each party performed a single measurement on his subsystem, only a finite amount of randomness could be certified.  In \cite{Ubounded17}, the authors proved one could certify any amount of random bits from a pair of qubits in a pure state when sequences of measurements were applied to each local system. Moreover, in the tripartite systems, the authors \cite{Mermin18,PRX21} studied the randomness by the Mermin-Ardehali-Belinskii-Klyshko (MABK) inequality and gave upper bounds on the amount of the randonmess. However, based on sequential measurement scenarios, whether unbounded  certifiable randomness can be obtained from a tripartite genuinely entangled state is not known. So our current work represents a step towards a better understanding of the limitations on how much device-independent randomness could be robustly generated from the multipartite entangled states.

It is also interesting to consider similar problem under the setting with multiple Alices, Bobs and Charlies.
Moreover,  one finds that our method do not help when considering the setting of Svetlichny nonlocality. Maybe there exists some state and measurement strategies such that more than 2 Charlies could share the Svetlichny nonlocality with single Alice and Bob. Moreover, it is unknown whether can we obtain an  unbounded sequential   violations of Mermin inequality or NS inequality with initial state being $|\mathrm{W}\rangle$ as  we only consider some special measurement strategies in our paper.

\bigskip
\noindent{\bf Acknowledgments}\, \,  This work is supported by  the National Natural Science Foundation of China (Grant No.11725417, No.11974057), NSAF(Grant No. U1930403), and Science Challenge Project(Grant No.2018005), National Natural Science Foundation of China (Grant No. 12005092), the China Postdoctoral Science Foundation (2020M681996), the Key Research and Development Project of Guangdong province under Grant No.2020B0303300001, the Guangdong Basic and Applied Research Foundation under Grant No.2020B1515310016, Key Lab of Guangzhou for Quantum Precision Measurement under Grant No.202201000010. Ya Xi and Mao-Sheng Li contribute equally to this work.


\onecolumngrid

\appendix
\section*{\textsc{Appendix}}
\section{The calculation of $I_M^{(k)}$}\label{appen:I_m}
Now we derive the Mermin value for the given measurement strategy. Let the measurement strategy of Alice be defined by the POVM effects

\begin{eqnarray}
  A_{0|0} = \frac{\mathbb{I}+\sigma_{1}}{2}, \ \   A_{0|1} =\frac{\mathbb{I}+\sigma_{2}}{2},
\end{eqnarray}
Bob's POVMs are defined by
\begin{eqnarray}
  B_{0|0} = \frac{\mathbb{I}-\theta\sigma_{2}}{2}, \ \
  B_{0|1} = \frac{\mathbb{I}+\theta\sigma_{1}}{2},
\end{eqnarray}
for $\theta\in[0,1]$. For each $k=1,2,\cdots,n$,  Charlie$^{(k)}$'s POVMs are defined by
\begin{eqnarray}
  C_{0|0}^{(k)} = \frac{\mathbb{I}+\sigma_{1}}{2}, \ \
  C_{0|1}^{(k)}= \frac{\mathbb{I}+\gamma_{k}\sigma_{2}}{2}.
\end{eqnarray}

The observables are given by $X_{i}= A_{0|i}-A_{1|i}$, $Y_{i}= B_{0|i}-B_{1|i}$, $Z^{(k)}_{i}= C^{(k)}_{0|i}-C^{(k)}_{1|i}$, $i=0,1.$

Let $\rho_{ABC^{(k-1)}}$ shared by Alice, Bob and Charlie$^{(k-1)}$ prior to Charlie$^{(k-1)}$'s measurements. Using the L\"{u}ders rule, the state is sent to Charlie$^{(k)}$ is
$$\begin{array}{rl}
   \rho_{ABC^{(k)}} & = \frac{1}{2}\sum\limits_{c,z}(\mathbb{I}\otimes\mathbb{I}\otimes\sqrt{C^{(k-1)}_{c|z}}\rho_{ABC^{(k-1)}}\mathbb{I}\otimes\mathbb{I}\otimes\sqrt{C^{(k-1)}_{c|z}})\\[3mm]
   & =\frac{1}{2}(\mathbb{I}\otimes\mathbb{I}\otimes\frac{\mathbb{I}+\sigma_{1}}{2}\rho_{ABC^{(k-1)}}\mathbb{I}\otimes\mathbb{I}\otimes\frac{\mathbb{I}+\sigma_{1}}{2}+\mathbb{I}\otimes\mathbb{I}\otimes\frac{\mathbb{I}-\sigma_{1}}{2}\rho_{ABC^{(k-1)}}\mathbb{I}\otimes\mathbb{I}\otimes\frac{\mathbb{I}-\sigma_{1}}{2}\\[3mm]
&+\mathbb{I}\otimes\mathbb{I}\otimes\sqrt{\frac{\mathbb{I}+\gamma_{k-1}\sigma_{2}}{2}}\rho_{ABC^{(k-1)}}\mathbb{I}\otimes\mathbb{I}\otimes\sqrt{\frac{\mathbb{I}+\gamma_{k-1}\sigma_{2}}{2}}+ \mathbb{I}\otimes\mathbb{I}\otimes\sqrt{\frac{\mathbb{I}-\gamma_{k-1}\sigma_{2}}{2}}\rho_{ABC^{(k-1)}}\mathbb{I}\otimes\mathbb{I}\otimes\sqrt{\frac{\mathbb{I}-\gamma_{k-1}\sigma_{2}}{2}} ) \\[3mm]
& =\frac{2+\sqrt{1-\gamma_{k-1}^{2}}}{4}\rho_{ABC^{(k-1)}}+\frac{1}{4}(\mathbb{I}\otimes\mathbb{I}\otimes\sigma_{1})\rho_{ABC^{(k-1)}}(\mathbb{I}\otimes\mathbb{I}\otimes\sigma_{1})\\[3mm]
&+\frac{1-\sqrt{1-\gamma_{k-1}^{2}}}{4}(\mathbb{I}\otimes\mathbb{I}\otimes\sigma_{2})\rho_{ABC^{(k-1)}}(\mathbb{I}\otimes\mathbb{I}\otimes\sigma_{2}),
\end{array}$$
where we use the identity for the final calculation
\begin{equation}
\sqrt{\frac{\mathbb{I}\pm\gamma_{k}\sigma_{\vec{r}}}{2}}=\frac{(\sqrt{1+\gamma_{k}}+\sqrt{1-\gamma_{k}})\mathbb{I}\pm(\sqrt{1+\gamma_{k}}-\sqrt{1-\gamma_{k}})\sigma_{\vec{r}}}{2\sqrt{2}}.
\end{equation}
Then we will consider the Mermin value of $\rho_{ABC^{(k)}}$:
$$\begin{array}{rl}
  \mathbf{I}^{(k)}_{M} & =\mathrm{Tr}[\rho_{ABC^{(k)}}(X_{1}Y_{0}Z^{(k)}_{0}+X_{0}Y_{1}Z^{(k)}_{0}+X_{0}Y_{0}Z^{(k)}_{1}-X_{1}Y_{1}Z^{(k)}_{1})]\\[4mm]
   & =\mathrm{Tr}[\rho_{ABC^{(k)}}(-\theta\sigma_{2}\otimes \sigma_{2}\otimes\sigma_{1}+\theta\sigma_{1}\otimes\sigma_{1}\otimes\sigma_{1}-\theta\gamma_{k}\sigma_{1}\otimes\sigma_{2}\otimes\sigma_{2}-\theta\gamma_{k}\sigma_{2}\otimes\sigma_{1}\otimes\sigma_{2})]\\[4mm]
&=2^{2-k}\theta[\gamma_{k}+\prod\limits_{j=1}^{k-1}(1+\sqrt{1-\gamma_{j}^{2}})].
\end{array}$$
In particular, $\mathbf{I}^{(1)}_{M}=2\theta(1+\gamma_{1}).$

\section{The calculation of $\mathbf{I}^{(k)}_{NS}$}\label{appen:I_NS}
Now we derive the $NS$ value for the given measurement strategy in the main text. Let the measurement strategy of Alice be defined by the POVM effects
\begin{eqnarray}
  A_{0|0} = \frac{\mathbb{I}+\cos(\theta)\sigma_{3}-\sin(\theta)\sigma_{1}}{2}, \ \
  A_{0|1}= \frac{\mathbb{I}+\cos(\theta)\sigma_{3}+\sin(\theta)\sigma_{1}}{2}.
\end{eqnarray}
Bob's POVMs are defined by
\begin{eqnarray}
  B_{0|0} = \frac{\mathbb{I}+\sigma_{3}}{2}, \ \
  B_{0|1} = \frac{\mathbb{I}+\sigma_{1}}{2}.
\end{eqnarray}
For each $k=1,2,\cdots,n$, Charlie$^{(k)}$'s POVMs are defined by
\begin{eqnarray}
  C_{0|0}^{(k)} = \frac{\mathbb{I}+\sigma_{3}}{2}, \ \
  C_{0|1}^{(k)} = \frac{\mathbb{I}+\gamma_{k}\sigma_{1}}{2}.
\end{eqnarray}
The observables are given by $X_{i}= A_{0|i}-A_{1|i}$, $Y_{i}= B_{0|i}-B_{1|i}$, $Z_{i}^{(k)}= C_{0|i}^{(k)}-C_{1|i}^{(k)}$, $i=0,1.$

Let $\rho_{ABC^{(k-1)}}$ shared by Alice, Bob and Charlie$^{(k-1)}$ prior to Charlie$^{(k-1)}$'s measurements. Using the L\"{u}ders rule, the state is sent to Charlie$^{(k)}$ is

$$\begin{array}{rl} \label{}
   \rho_{ABC^{(k)}} & = \frac{1}{2}\sum\limits_{c,z}(\mathbb{I}\otimes\mathbb{I}\otimes\sqrt{C^{(k-1)}_{c|z}}\rho_{ABC^{(k-1)}}\mathbb{I}\otimes\mathbb{I}\otimes\sqrt{C^{(k-1)}_{c|z}})\\[4mm]
   & =\frac{1}{2}(\mathbb{I}\otimes\mathbb{I}\otimes\frac{\mathbb{I}+\sigma_{3}}{2}\rho_{ABC^{(k-1)}}\mathbb{I}\otimes\mathbb{I}\otimes\frac{\mathbb{I}+\sigma_{3}}{2}+\mathbb{I}\otimes\mathbb{I}\otimes\frac{\mathbb{I}-\sigma_{3}}{2}\rho_{ABC^{(k-1)}}\mathbb{I}\otimes\mathbb{I}\otimes\frac{\mathbb{I}-\sigma_{3}}{2}\\[4mm]
&+\mathbb{I}\otimes\mathbb{I}\otimes\sqrt{\frac{\mathbb{I}+\gamma_{k-1}\sigma_{1}}{2}}\rho_{ABC^{(k-1)}}\mathbb{I}\otimes\mathbb{I}\otimes\sqrt{\frac{\mathbb{I}+\gamma_{k-1}\sigma_{1}}{2}}+ \mathbb{I}\otimes\mathbb{I}\otimes\sqrt{\frac{\mathbb{I}-\gamma_{k-1}\sigma_{1}}{2}}\rho_{ABC^{(k-1)}}\mathbb{I}\otimes\mathbb{I}\otimes\sqrt{\frac{\mathbb{I}-\gamma_{k-1}\sigma_{1}}{2}} ) \\[4mm]
& =\frac{2+\sqrt{1-\gamma_{k-1}^{2}}}{4}\rho_{ABC^{(k-1)}}+\frac{1}{4}(\mathbb{I}\otimes\mathbb{I}\otimes\sigma_{3})\rho_{ABC^{(k-1)}}(\mathbb{I}\otimes\mathbb{I}\otimes\sigma_{3})\\[4mm]
&+\frac{1-\sqrt{1-\gamma_{k-1}^{2}}}{4}(\mathbb{I}\otimes\mathbb{I}\otimes\sigma_{1})\rho_{ABC^{(k-1)}}(\mathbb{I}\otimes\mathbb{I}\otimes\sigma_{1}),
\end{array}$$
where we use the identity for the final calculation
\begin{equation}
\sqrt{\frac{\mathbb{I}\pm\gamma_{k}\sigma_{\vec{r}}}{2}}=\frac{(\sqrt{1+\gamma_{k}}+\sqrt{1-\gamma_{k}})\mathbb{I}\pm(\sqrt{1+\gamma_{k}}-\sqrt{1-\gamma_{k}})\sigma_{\vec{r}}}{2\sqrt{2}}.
\end{equation}
Then we will consider the $NS$ value of $\rho_{A^{(k)}BC}$:
$$\begin{array}{rl}
  \mathbf{I}^{(k)}_{NS} & =\mathrm{Tr}[\rho_{ABC^{(k)}}( Y_{0}Z_{0}^{(k)}+X_{0}Z_{0}^{(k)}
 + X_{1}Y_{0}- X_{0}Y_{1}Z_{1}^{(k)}+X_{1}Y_{1}Z_{1}^{(k)})]\\[3mm]
   & =\mathrm{Tr}[\rho_{ABC^{(k)}}(\mathbb{I}\otimes\sigma_{3}\otimes\sigma_{3}+[\cos(\theta)\sigma_{3}-\sin(\theta)\sigma_{1}]\otimes\mathbb{I}\otimes\sigma_{3}+[\cos(\theta)\sigma_{3}+\sin(\theta)\sigma_{1}]\otimes\sigma_{3}\otimes\mathbb{I}\\[3mm]
&-\gamma_{k}[\cos(\theta)\sigma_{3}-\sin(\theta)\sigma_{1}]\otimes\sigma_{1}\otimes\sigma_{1}+\gamma_{k}[\cos(\theta)\sigma_{3}+\sin(\theta)\sigma_{1}]\otimes\sigma_{1}\otimes\sigma_{1}]\\[3mm]
&=2^{2-k}\gamma_{k}\sin(\theta)+\frac{\prod\limits_{j=1}^{k-1}(1+\sqrt{1-\gamma_{j}^{2}})}{2^{k-1}}[1+\cos(\theta)]+\cos(\theta).
\end{array}$$
In particular, $\mathbf{I}^{(1)}_{NS}=1+2\cos(\theta)+2\gamma_{1}\sin(\theta).$

\section{The Proof of  Theorem \ref{theorem2}}\label{appen:TheoremI_m}
For the given measurements in the main text, in order to observe $\mathbf{I}^{(k)}_{M}>2$, we will need
\begin{equation}
 \mathbf{I}^{(k)}_{M}>2\Leftrightarrow \gamma_{k}>\frac{2^{k-1}}{\theta}-\prod\limits_{j=1}^{k-1}(1+\sqrt{1-\gamma_{j}^{2}}).
\end{equation}
Next we can define $\{\gamma_{k}(\theta)\}$, for some fixed $\epsilon>0$,
\begin{equation}\label{}
  \gamma_{k}(\theta)=\left\{
                       \begin{array}{ll}
                         (1+\epsilon)(\frac{1}{\theta}-1), & \hbox{$k=1$;} \\[2mm]
                     (1+\epsilon)(\frac{2^{k-1}}{\theta}-P_{k}) , & \hbox{$0\leq\gamma_{k-1}(\theta)\leq1$ ;} \\[2mm]
                         \infty, & \hbox{others,}
                       \end{array}
                     \right.
\end{equation}
where $P_{k}=\prod\limits_{j=1}^{k-1}(1+\sqrt{1-\gamma_{j}^{2}}).$

Then we can get
\begin{equation}
  \frac{\gamma_{k}(\theta)}{\gamma_{k-1}(\theta)}>2\Leftrightarrow 0<\gamma_{k-1}(\theta)\leq1.
\end{equation}
Here $\gamma_{1}(\theta)=(1+\epsilon)(\frac{1}{\theta}-1)$ and $\lim\limits_{\theta\rightarrow 1^{-}}\gamma_{1}(\theta)=0.$

By the induction, we can suppose there exists a $\theta_{k-1}$ such that on the interval $ (\theta_{k-1},1)$, all $\gamma_{i}(\theta)\in(0,1),$  and $\lim\limits_{\theta\rightarrow 1^{-}}\gamma_{i}(\theta)=0$  for $i=1,2,\cdots ,k-1.$ Then according to the definition of $\gamma_{k}(\theta)$, we will have $$\lim\limits_{\theta\rightarrow 1^{-}}\gamma_{k}(\theta)=\lim\limits_{\theta\rightarrow 1^{-}} (1+\epsilon)(\frac{2^{k-1}}{\theta}-P_{k}) =(1+\epsilon)(2^{k-1}-2^{k-1})=0,$$
here we use the limit $\lim\limits_{\theta\rightarrow 1^{-}} P_k=2^{k-1} $ which holds as the induction assumptions $\lim\limits_{\theta\rightarrow 1^{-}}\gamma_{i}(\theta)=0$  for $i=1,2,\cdots ,k-1.$  So  $\forall n\in\mathbb{N},$ we can find a $\theta_{n}\in(0,1)$ such that $0<\gamma_{1}(\theta)<\gamma_{2}(\theta)<\cdots<\gamma_{n}(\theta)<1$ for all $\theta\in(\theta_{n},1).$

\section{The Proof of  {\it Remark 2} }\label{appen:Remark 2}
Alice's POVMs are defined by
\begin{eqnarray}
	A_{0|0}  =  \frac{\mathbb{I}+\cos(\theta_{1})\sigma_{3}-\sin(\theta_{1})\sigma_{1}}{2}, \ \ \
	A_{0|1}  =  \frac{\mathbb{I}+\sin(\theta_{1})\sigma_{3}+\cos(\theta_{1})\sigma_{1}}{2}.
\end{eqnarray}
Bob's POVMs are defined by
\begin{eqnarray}
	B_{0|0}  =  \frac{\mathbb{I}+\cos(\theta_{2})\sigma_{3}-\sin(\theta_{2})\sigma_{1}}{2}, \ \ \ \
	B_{0|1}  =  \frac{\mathbb{I}+\sin(\theta_{2})\sigma_{3}+\cos(\theta_{2})\sigma_{1}}{2},
\end{eqnarray}
for $\theta_{i}\in[0,\frac{\pi}{2}], i\in\{1,2\}$. For each $k=1,2,\cdots,n$,  Charlie$^{(k)}$'s POVMs are defined by
\begin{eqnarray}
  C_{0|0}^{(k)} = \frac{\mathbb{I}-\sigma_{3}}{2}, \ \
  C_{0|1}^{(k)} = \frac{\mathbb{I}-\gamma_{k}\sigma_{1}}{2}.
\end{eqnarray}

Under these measurements  and   the initial state $|\mathrm{W}\rangle=\frac{1}{\sqrt{3}}(|100\rangle+|010\rangle+|001\rangle),$ we can get
\begin{equation}
\mathbf{I}^{(k)}_{M}=2^{1-k}[\frac{5}{3}\sin(\theta_{1}+\theta_{2})\prod\limits_{j=1}^{k-1}(1+\sqrt{1-\gamma_{j}^{2}})+\frac{4}{3}\sin(\theta_{1}+\theta_{2})\gamma_{k}].
\end{equation}
Note that
\begin{equation}
 \mathbf{I}^{(k)}_{M}>2\Leftrightarrow \gamma_{k}>\frac{2^{k}-\frac{5}{3}\sin(\theta_{1}+\theta_{2})\prod\limits_{j=1}^{k-1}(1+\sqrt{1-\gamma_{j}^{2}})}{\frac{4}{3}\sin(\theta_{1}+\theta_{2})}.
\end{equation}
Next we can define $\{\gamma_{k}(\theta_{1}, \theta_{2})\}$, for some fixed $\epsilon>0$,
\begin{equation}
  \gamma_{k}(\theta_{1},\theta_{2} )=\left\{
                       \begin{array}{ll}
                         (1+\epsilon)\frac{2-\frac{5}{3}\sin(\theta_{1}+\theta_{2})}{\frac{4}{3}\sin(\theta_{1}+\theta_{2})}, & \hbox{$k=1$;} \\[2mm]
                     (1+\epsilon)\frac{2^{k}-\frac{5}{3}\sin(\theta_{1}+\theta_{2})P_{k}}{\frac{4}{3}\sin(\theta_{1}+\theta_{2})} , & \hbox{$0\leq\gamma_{k-1}(\theta_{1}, \theta_{2})\leq1$ ;} \\
                         \infty, & \hbox{others,}
                       \end{array}
                     \right.
\end{equation}
where $P_{k}=\prod\limits_{j=1}^{k-1}(1+\sqrt{1-\gamma_{j}^{2}}).$ Then we can get
\begin{equation}\label{eq:D7}
  \frac{\gamma_{k}(\theta_{1},\theta_{2})}{\gamma_{k-1}(\theta_{1},\theta_{2})}>2\Leftrightarrow 0<\gamma_{k-1}(\theta_{1},\theta_{2})\leq1.
\end{equation}
Note that
$$  \gamma_{1}(\theta_{1},\theta_{2})=(1+\epsilon)\frac{2-\frac{5}{3}\sin(\theta_{1}+\theta_{2})}{\frac{4}{3}\sin(\theta_{1}+\theta_{2})}=(1+\epsilon)[\frac{6}{4\sin (\theta_1+\theta_2)} -\frac{5}{4}]\geq \frac{1+\epsilon}{4}.$$ By Eq. \eqref{eq:D7}, we would have $\gamma_{2}(\theta_{1},\theta_{2})>\frac{1}{2}(1+\epsilon)$ and $\gamma_{3}(\theta_{1},\theta_{2})>1+\epsilon.$ Therefore,  in the above strategy, at most two Charlies can demonstrate standard nonlocality through the violation of Mermin inequality with the single Alice and Bob.

\section{The Proof  of Theorem \ref{theorem4}}\label{appen:TheoremI_NS}
For the given measurements in the main text, in order to observe $\mathbf{I}^{(k)}_{NS}>3$, we will need
\begin{equation}
 \mathbf{I}^{(k)}_{NS}>3\Leftrightarrow \gamma_{k}>\frac{3-\cos(\theta)-(1+\cos(\theta))\frac{\prod\limits_{j=1}^{k-1}(1+\sqrt{1-\gamma_{j}^{2}})}{2^{k-1}}}{2^{2-k}\sin(\theta)}.
\end{equation}
Next we can define $\{\gamma_{k}(\theta)\}$, for some fixed $\epsilon>0$,
\begin{equation}\label{}
  \gamma_{k}(\theta)=\left\{
                       \begin{array}{ll}
                         (1+\epsilon)\frac{1-\cos(\theta)}{\sin(\theta)}, & \hbox{k=1;} \\
                     (1+\epsilon)\frac{3-\cos(\theta)-(1+\cos(\theta))\frac{P_{k}}{2^{k-1}}}{2^{2-k}\sin(\theta)} , & \hbox{$0\leq\gamma_{k-1}(\theta)\leq1$ ;} \\
                         \infty, & \hbox{others,}
                       \end{array}
                     \right.
\end{equation}
where $P_{k}=\prod\limits_{j=1}^{k-1}(1+\sqrt{1-\gamma_{j}^{2}}).$
Then we can get
\begin{equation}
  \frac{\gamma_{k}(\theta)}{\gamma_{k-1}(\theta)}>2\Leftrightarrow 0<\gamma_{k-1}(\theta)\leq1,
\end{equation}
here $\gamma_{1}(\theta)=(1+\epsilon)\frac{1-\cos(\theta)}{\sin(\theta)}$ and $\lim\limits_{\theta\rightarrow 0^{+}}\gamma_{1}(\theta)=0.$

By the induction, we can suppose there exists a $\theta_{k-1}$ such that on the interval $(0,\theta_{k-1})$, all $\gamma_{i}(\theta)\in(0,1),$  and $\lim\limits_{\theta\rightarrow 0^{+}}\gamma_{i}(\theta)=0$  for $i=1,2,\cdots ,k-1.$ Note that when looking $P_k$ as a function on the small interval $(0,\theta_{k-1})$, its  differetial could be calculated as
$$
 P'_k(\theta)=\sum_{j=1}^{k-1}\left( \frac{-2 \gamma_j \gamma_j'}{2\sqrt{1-\gamma_j^2}}\right) \frac{P_k}{1+\sqrt{1-\gamma_j^2}}
$$
which is tending to $0$ as $\theta\rightarrow 0^{+}.$
 Then according to the definition of $\gamma_{k}(\theta)$, by L'Hopspital rule,  we will have $$\lim\limits_{\theta\rightarrow 0^{+}}\gamma_{k}(\theta)=\lim\limits_{\theta\rightarrow 0^{+}} (1+\epsilon)\frac{ \sin \theta-\frac{-\sin \theta P_k(\theta)+(1+\cos \theta ) P_k'(\theta)  }{2^{k-1}}}{2^{2-k}\cos(\theta)}=0.$$
   So  $\forall n\in\mathbb{N},$ we can find a $\theta_{n}\in(0,1)$ such that $0<\gamma_{1}(\theta)<\gamma_{2}(\theta)<\cdots<\gamma_{n}(\theta)<1$ for all $\theta\in(0,\theta_{n}).$

\section{The Proof of  {\it Remark 3} }\label{appen:Remark 3}

Alice's POVMs are defined by
\begin{eqnarray}
  A_{0|0} =\frac{\mathbb{I}+\sigma_{3}}{2}, \ \
  A_{0|1} = \frac{\mathbb{I}+\sigma_{1}}{2}.
\end{eqnarray}
Bob's POVMs are defined by
\begin{eqnarray}
  B_{0|0} = \frac{\mathbb{I}-\sigma_{3}}{2}, \ \
  B_{0|1} = \frac{\mathbb{I}+\sigma_{1}}{2}.
\end{eqnarray}
For each $k=1,2,\cdots,n$,  Charlie$^{(k)}$'s POVMs are defined by
\begin{eqnarray}
  C_{0|0}^{(k)} = \frac{\mathbb{I}+\sigma_{3}}{2}, \ \
  C_{0|1}^{(k)} = \frac{\mathbb{I}+\gamma_{k}\sigma_{1}}{2}.
\end{eqnarray}
Moreover,  the corresponding $NS$ inequality is :
\begin{equation}
	\langle X_{1}Y_{1}\rangle+\langle Y_{0}Z_{0}\rangle+\langle X_{1} Z_{1}\rangle+\langle X_{0}Y_{0}Z_{0}\rangle-\langle X_{1}Y_{0}Z_{1}\rangle\leq3.
\end{equation}
Under these measurements  and   the initial state $|\mathrm{W}\rangle=\frac{1}{\sqrt{3}}(|100\rangle+|010\rangle+|001\rangle),$ we can get
\begin{equation}
\mathbf{I}^{(k)}_{NS}=\frac{2}{3}+\frac{2^{3-k}}{3}(\gamma_{k}+\prod\limits_{j=1}^{k-1}(1+\sqrt{1-\gamma_{j}^{2}})).
\end{equation}
Therefore, we have
\begin{equation}
 \mathbf{I}^{(k)}_{NS}>2\Leftrightarrow \gamma_{k}>\frac{7}{2^{3-k}}-\prod\limits_{j=1}^{k-1}(1+\sqrt{1-\gamma_{j}^{2}}).
\end{equation}
Next we can define $\{\gamma_{k}\}$, for some fixed $\epsilon>0$,
\begin{equation}\label{}
  \gamma_{k}=\left\{
                       \begin{array}{ll}
                         \frac{3}{4}(1+\epsilon), & \hbox{$k=1$;} \\[2mm]
                     (1+\epsilon)(\frac{7}{2^{3-k}}-P_{k}) , & \hbox{$0\leq\gamma_{k-1}\leq1$ ;} \\[2mm]
                         \infty, & \hbox{others.}
                       \end{array}
                     \right.
\end{equation}
where $P_{k}=\prod\limits_{j=1}^{k-1}(1+\sqrt{1-\gamma_{j}^{2}}).$

Then we can get
\begin{equation}
  \frac{\gamma_{k}}{\gamma_{k-1}}>2\Leftrightarrow 0<\gamma_{k-1}\leq1.
\end{equation}
Here $\gamma_{1}=\frac{3}{4}(1+\epsilon),$ then $\gamma_{2}>2\gamma_{1}=\frac{3}{2}(1+\epsilon)>1.$ So  in our setting, at most one Charlie can demonstrate genuinely nonsignal nonlocality through the violation of $NS$ inequality with the single Alice and Bob.

\end{document}